\documentstyle[12pt]{article}
\newcommand{\bq}{\begin{equation}}
\newcommand{\eq}{\end{equation}}
\newcommand{\ba}{\begin{eqnarray}}
\newcommand{\ea}{\end{eqnarray}}

\newcommand{\fr}[2]{\frac{#1}{#2}}
\newcommand{\non}{\nonumber}

\newcommand{\al}{\mbox{$\alpha$}}
\newcommand{\om}{\mbox{$\omega$}}
\newcommand{\en}{\mbox{$\epsilon_n$}}
\newcommand{\ep}{\mbox{${\cal E}_n$}}
\newcommand{\de}{\partial}
\newcommand{\cb}{\mbox{$\frac{1}{r^2}$}}
\newcommand{\qu}{\mbox{$\frac{1}{r}$}}
\newcommand{\la}{\left\langle}
\newcommand{\ra}{\right\rangle}
\newcommand{\r}{\mbox{$\vec{r}$}}
\newcommand{\n}{\mbox{$\vec{n}$}}
\newcommand{\lv}{\mbox{$\vec{L}$}}
\newcommand{\pv}{\mbox{$\vec{p}$}}
\newcommand{\sv}{\mbox{$\vec{\sigma}$}}

\newcommand{\pp}{\mbox{$\vec{p}\,'$}}
\newcommand{\qv}{\mbox{$\vec{q}$}}
\newcommand{\kv}{\mbox{$\vec{k}$}}
\newcommand{\kp}{\mbox{$\vec{k}\,'$}}

\newcommand{\Sv}{\mbox{$\vec{\cal S}$}}
\newcommand{\pop}[1]{\mbox{$\lambda_+(#1)$}}

\begin{document}
\pagestyle{empty}

\hfill BINP--93--80

\hfill September 1993
\vspace{1.0 cm}
\begin{center}{\Large \bf Order $\alpha^4 R_{\infty}$ corrections to
positronium $P$ levels}\\

\vspace{1.0 cm}

{\bf  I.B. Khriplovich, A.I. Milstein, and A.S. Yelkhovsky} \\
\vspace{0.5 cm}
Budker Institute of Nuclear Physics, 630090 Novosibirsk, Russia
\end{center}

\vspace{4.0 cm}

\begin{abstract}
The order $\alpha^4 R_{\infty}$ corrections to positronium $P$ levels are
found. The calculation is reduced to the ordinary perturbation theory for
the nonrelativistic Schr\"odinger equation.  The perturbation operators have
the Breit-type structure and are obtained by calculating on-mass-shell
diagrams. Found energy corrections  constitute numerically $\delta E(2 ^1
P_1)=0.06$ MHz, $\delta E(2 ^3 P_2)=0.08$ MHz, $\delta E(2 ^3 P_1)=0.025$
MHz, $\delta E(2 ^3 P_2)=-0.58$ MHz.

\end{abstract}

\newpage
\pagestyle{plain}
\pagenumbering{arabic}

\section{Introduction}

Precision measurements of positronium structure provide a unique test of
quantum electrodynamics. The typical accuracy reached in the measurements of
the positronium $2 ^3 S_1 - 2 ^3 P_J (J=0,1,2)$, \ $1 ^3 S_1 - 2 ^3 S_1$ and $2
^3 S_1 - 2 ^1 P_1$ intervals constitutes few MHz
\cite{mbc,hcr,dfc,fmc,chl,hage,hag2}. The order $\al^3 \log(1/\al)
R_{\infty}$ and $\al^3 R_{\infty}$ corrections
($R_{\infty}=109\,737.315\,682\,7(48)$ cm$^{-1}$ is the Rydberg constant)
\cite{fm,ful,grs} are insufficient now for the comparison of quantum
electrodynamics with those experimental data.

The two-body bound state QED problem is by itself of independent theoretical
interest. The generally accepted theoretical approach to it goes back to
Refs \cite{gr,le,br}. This approach starts from the introduction of a
relativistic two-body wave equation, which can be solved exactly and in the
nonrelativistic limit reduces to the Schr\"odinger equation. Then a
perturbation series is developed about the exact solution.

Our approach is different and is described, as applied to the corrections
logarithmic in \al , in Refs \cite{ky1,kmy1,kmy}. The corrections discussed
in Refs \cite{ky1,kmy1,kmy} and here, are of relativistic origin and can be
found as follows. We construct effective perturbation operators by
calculating on-mass-shell diagrams and then use those operators in the
standard perturbation theory for the nonrelativistic Schr\"odinger
equation.

The order $\al^4\log(1/\al) R_{\infty}$ corrections to the
positronium levels were calculated recently by Fell \cite{fel} and then by
us \cite{kmy} (we have corrected in Ref \cite{kmy} a numerical error made
in Ref \cite{kmy1}). This shift exists in the $S$ states only and scales with
the principal quantum number $n$ as $n^{-3}$. The logarithmic structure of
this correction allows one to treat the relativistic effects as a
perturbation when deriving the result.  However, if one tries to go beyond
the logarithmic approximation, the logarithmic integrals which are cut off
at the electron mass $m$ should be treated exactly, and the problem becomes
extremely tedious.

Fortunately, for states of higher angular momenta, $L>0$, the situation is
better since their nonrelativistic wave functions fall off at
small distances. Therefore, the integrals arising in the perturbation
theory, converge in the nonrelativistic region which makes the problem quite
tractable. The main complication (underestimated by us at the beginning of
the work) is of the ``book-keeping'' nature.

Here we present the results of our analytical calculations for the order
$\alpha^4 R_{\infty}$ corrections to positronium $P$ levels. In the case
$n=2$ the results for the fine-splitting of $P$ levels can be directly
compared with the data extracted from the experimental results of Refs
\cite{hcr,chl,hage,hag2}.

Analogous corrections for the electron-electron
interaction in helium were obtained numerically in Ref \cite{dk}.

\section{Contributions of irreducible operators}

\subsection{Relativistic correction to the dispersion law}

Let us start with the kinematic correction generated by the $v^4/c^4$ term
in the dispersion law for electron and positron,
\ba
\sqrt{m^2 + p^2} - m &=& \fr{p^2}{2m} - \fr{p^4}{8m^3} + \fr{p^6}{16m^5} +
                         \ldots , \\
E^{(1)}_{kin} &=& 2 \la \fr{p^6}{16m^5} \ra .
\ea
By means of motion equations we transform the expectation value to
\ba
E^{(1)}_{kin} &=& \fr{m\al^6}{64} \la (\ep +\qu)\fr{p^2}{2}(\ep +\qu) \ra
                  \non \\
              &=& \fr{m\al^6}{64} \la \ep^3 + \fr{3\ep^2}{r} +
                  \fr{3\ep}{r^2} + \fr{1}{r^3} + \fr{1}{2r^4} \ra; \;\;\;
                  \ep \equiv \fr{2E_n}{m\al^2} = - \fr{1}{2n^2}.
\ea
After extracting the overall factor $m\al^6$ everything else here and below
is written in the usual atomic units. The substitution of the
nonrelativistic Coulomb expectation values for $1/r^k$,
\ba\label{eq:means}
\la r^{-1} \ra =  \fr{1}{n^2}, &&\;\;\; \la r^{-2} \ra =
                   \fr{2}{3n^3}, \\
\la r^{-3} \ra = \fr{1}{3n^3}, &&\;\;\; \la r^{-4} \ra =
                   \fr{2}{5n^3}\left( 1 - \fr{2}{3n^2} \right),
\ea
reduces this energy correction to
\bq
E^{(1)}_{kin} = \fr{\en}{2^6\cdot 3 \cdot 5}\left( 8 - \fr{17}{n^2} +
                \fr{75}{8n^3} \right),
\eq
where $\en \equiv m\al^6/n^3$.

\subsection{Relativistic corrections to the Coulomb interaction}

This perturbation operator will be extracted from the scattering amplitude
for free particles. It is convenient to consider the positron as electron of
opposite charge. Then the scattering amplitude due to the single
Coulomb exchange is
\bq\label{eq:AC}
A_C = - \fr{4\pi\al}{q^2} \rho (\pp, \pv) \rho (-\pp, -\pv),
\eq
where
\bq
\rho (\pp, \pv) = u^+(\pp) u(\pv), \;\;\; \qv = \pp - \pv.
\eq
We define scattering amplitudes with the sign opposite to the standard
one, so that it can be immediately identified with the matrix element of an
interaction operator in the momentum representation.  Let us substitute into
(\ref{eq:AC}) the solution of the free Dirac equation
\bq\label{eq:u}
u(\pv) = \sqrt{\fr{2\om_{\vec{p}}}{\om_{\vec{p}}+m}} \pop{\pv} w.
\eq
where $w$ is a bispinor describing particle at rest,
\[\lambda_{+}(\vec{p})=\frac{1}{2}\left(1+\frac{\vec{\alpha}\vec{p}+\beta m}
{\omega_{\vec{p}}}\right),\;\; \om_{\vec{p}} = \sqrt{p^2 + m^2}.\]
The corrections of fourth order in $v/c$ are
\bq\label{eq:C+}
V_{C} = - \fr{\al}{16m^4} \fr{4\pi}{q^2} \left\{ (p'^2-p^2)^2 -
i(\Sv,\pp\times\pv) (q^2+3p'^2+3p^2) - 2(\Sv,\pp\times\pv)^2 \right\}.
\eq
We have neglected here the operator proportional to $p^2+p'^2$ since its
expectation value in the coordinate representation $\la \Delta \delta(\r)
+ \delta(\r) \Delta \ra$ vanishes for $P$ states. The expectation value of
operator (\ref{eq:C+}) is again conveniently calculated in the coordinate
representation. Its spin-independent part is
\ba
\la \fr{4\pi}{q^2} (p'^2-p^2)^2 \ra &=& \la \left[ p^2, [p^2, \qu]\right]\ra
\\ &=& 2 \la \left[\qu, [p^2, \qu]\right] \ra = 4 \la r^{-4} \ra.
\ea
The treatment of the spin-dependent operators is somewhat more complicated.
Here we pass over to the Fourier transforms of the operators:
\ba
\int \fr{d^3q}{(2\pi)^3} \fr{4\pi \qv}{q^2}e^{i\qv\r} &=&
\fr{i\n}{r^2}, \label{eq:fu1}\\
\int \fr{d^3q}{(2\pi)^3} \fr{4\pi q_i q_j}{q^2}e^{i\qv\r} &=&
\fr{\delta_{ij} - 3n_i n_j}{r^3} + 4\pi n_i n_j \delta (\r),\label{eq:fu2}
\ea
then use motion equations and expectation values (\ref{eq:means}),
as well as the value of the radial wave function derivative at the origin
\bq
|R'_{n1}(0)|^2 = \frac{4}{9n^2} \left( 1 - \frac{1}{n^2} \right).
\eq
In this way we get
\ba
E^{(1)}_{C} &=& \fr{\en}{2^6\cdot 5^2} \left\{ -5 \left( 1 -
\fr{2}{3n^2} \right) - \Sv\lv \left( 19 - \fr{121}{6n^2} \right) + \fr{
(\Sv\lv)^2 + \Sv^2}{3} \left( 1 - \fr{3}{2n^2} \right) \right\}.
\ea

Now, due to the Coulomb interaction electron (positron) can go over into
a negative-energy intermediate state. The corresponding contributions are
described by $Z$-diagrams of the kind presented in Fig.1. Here the particle
staying in a positive-energy state can be treated in the nonrelativistic
approximation. The point is that the large energy denominator, equal
approximately to $2m \gg E_n$, due to "heavy" intermediate states, and small
matrix element of the $Z$-line provide the sufficient power of \al\ in the
perturbation. In this way we get for the perturbation operator
\ba
V_{C-} &=& \fr{(4\pi\al)^2}{2m} \int \fr{d^3 k}{(2\pi)^3} \fr{1}{k^2(\vec{k}
           - \qv)^2} \times \non \\
       &&  \left( \fr{1}{2}\left( 1 - \fr{m}{\om_{\vec{p}+\vec{k}}}
           \right) + \fr{\sv\pp}{2m}\fr{\sv\pv}{2m} - \fr{\sv\pp}{2m}
           \fr{\sv(\pv+\vec{k})}{2m} - \fr{\sv(\pv+\vec{k})}{2m}\fr{\sv\pv}{2m}
           \right) \non \\
&\rightarrow& - \fr{(4\pi\al)^2}{(2m)^3} \int \fr{d^3 k}{(2\pi)^3}
              \fr{\vec{k}(\qv-\vec{k})}{k^2(\qv - \vec{k})^2}.\label{eq:C-}
\ea
Going over to the coordinate representation, we note that the last
integral is in fact the convolution of the Fourier-transform of the operator
$i\n/r^2$ with itself. So,
\bq
E^{(1)}_{C-} = \fr{m\al^6}{2^6}\la r^{-4} \ra = \fr{\en}{2^5\cdot 5}
               \left( 1 - \fr{2}{3n^2} \right).
\eq
Let us note that actually the integral in (\ref{eq:C-}) diverges at large
$k$. It can be easily seen however that the divergent part is independent of
the momentum transfer \qv . Therefore in the coordinate representation the
corresponding operator is just $\delta(\r)$. Its expectation value does not
vanish in $S$ states only and leads there (via the accurate cut-off of the
linear divergence at $k\sim m$) to a correction of the order $m\al^5$.

\subsection{Single magnetic exchange}

In the noncovariant perturbation theory the electron-positron scattering
amplitude due to the exchange by one magnetic quantum is
\bq\label{eq:AM}
A_M = - \fr{4\pi\al}{q} \fr{j_i (\pp, \pv) j_j (-\pp, -\pv)}{E_n - q -
\fr{p^2+p'^2}{2m}}\left( \delta_{ij} - \fr{q_i q_j}{q^2} \right).
\eq
Here
\bq\label{eq:j}
\vec{j}(\pp, \pv) = u^+(\pp)\vec{\al} u(\pv)
\eq
is the matrix element of the current taken over the solutions (\ref{eq:u})
of the free Dirac equation. In the dispersion law for electron and positron it
is sufficient here to confine to the nonrelativistic approximation.

Let us start from the contribution to the perturbation operator produced
by the $v^{2}/c^{2}$ corrections to the currents:
\ba
V_{curr} &=& \fr{\al}{4m^4} \fr{4\pi}{q^2} \left\{ 4p'^2 \left( p^2 -
         \fr{(\pv\qv)^2}{q^2} + \pv\qv \fr{i(\qv\times\pv,\Sv)}{q^2} +
         \fr{1}{2}(\qv\Sv)^2 \right) \right. \non \\
         && \left. + \fr{p'^2-p^2}{2} \left( (p'^2-p^2)\left( 1 +
         \fr{i(\qv\times\pv,\Sv)}{q^2} - \Sv^2 \right) + (2\pv+\qv)\Sv\;
         (\qv\Sv) \right) \right\}.
\ea
Here we have neglected again the terms with vanishing $P$ state
expectation values. Going over to the coordinate representation by means of
(\ref{eq:fu1}), (\ref{eq:fu2}) and
\ba
\int \fr{d^3q}{(2\pi)^3} \fr{4\pi q_i q_j}{q^4}e^{i\qv\r} &=&
\fr{\delta_{ij} - n_i n_j}{2r}, \\
\int \fr{d^3q}{(2\pi)^3} \fr{4\pi \qv}{q^4}e^{i\qv\r} &=& \fr{i\n}{2},
\ea
we obtain
\bq
E^{(1)}_{curr} = \fr{\en}{2^5\cdot 5} \left\{ \fr{17}{3} -
                   \fr{11}{n^2} + \fr{5}{n^3} - \fr{7\Sv\lv}{2} \left( 1 -
                   \fr{1}{n^2} \right)
                  + 2(\Sv\lv)^2 \left( 1 - \fr{7}{6n^2} \right) -
                   2\Sv^2 \left( 1 - \fr{1}{n^2} \right) \right\}.
\eq

Let us consider now the retardation effect. To this end the currents can
be taken in the leading approximation:
\bq\label{eq:Pc}
\vec{j}(\pp, \pv) \rightarrow \fr{1}{2m} ( \pp + \pv + i\qv\times\sv ).
\eq
At the atomic momentum transfer, $q\sim m\al$, the perturbation of interest
originates from the second-order term of the expansion of the factor
$[E_n-(p^2+p'^2)/2m-q]^{-1}$ in (\ref{eq:AM}) in powers of
$(E_n-(p^2+p'^2)/2m)/q$:
\bq
V_{ret} = - \fr{\al}{2m^2} \fr{4\pi}{q^2} \fr{\left( E_n -
            \fr{p^2+p'^2}{2m} \right)^2}{q^2}
            \left\{ q^2 + 2\pv\pp - 2\fr{(\qv\pp)(\qv\pv)}{q^2} -
            2i(\pv\times\qv,\Sv) - q^2\Sv^2 + (\qv\Sv)^2 \right\}.
\eq
At first sight, the expectation value of this operator diverges linearly
at small $q$. This divergence can be demonstrated however to be unrelated
to the order $m\al^6$ correction we are interested in. Indeed, let us
split the region of integration over $q$ into two, those from 0 to $\lambda$
and from $\lambda$ to $\infty$ where $m\al^2 \ll \lambda \ll m\al$. In the
second region our expansion is applicable and the result of integration
contains a term proportional to $1/\lambda$. Since the initial integral is
independent of $\lambda$, this term cancels in the sum with the integral
over the first region calculated without the expansion. Meanwhile, this
last integral has no contribution of the $\al^4 R_{\infty}$ order
independent of $\lambda$.  Therefore, taking as the Fourier-transform of
\bq
\fr{4\pi}{q^4} \left( \pv\pp - \fr{(\qv\pp)(\qv\pv)}{q^2} \right),
\eq
the operator
\bq
- \fr{r}{8} ( 3p^2 - (\n\pv)^2 ),
\eq
we get finally:
\bq
E^{(1)}_{ret} = \fr{\en}{2^5\cdot 3 \cdot 5} \left( 14 - \fr{15}{n} +
                \fr{13}{2n^2} \right).
\eq

Magnetic quantum propagates for a finite time and can cross arbitrary number
of the Coulomb ones. Simple counting of the momenta powers demonstrates
that it is sufficient to include the diagrams with one and two Coulomb
quanta (dashed lines) crossed by the magnetic photon (wavy line). In the
first case, Fig.2, the perturbation operator arises as a product of the
Pauli currents (\ref{eq:Pc}) and the first-order term in the expansion in
$(E_n-(p^2+p'^2)/2m)/q$:
\ba\label{eq:MC}
V_{MC} &=& \fr{\al^2}{m^2} \int \fr{d^3 k}{(2\pi)^3}
           \fr{(4\pi)^2}{k^4 (\qv- \kv)^2}
           \left( \fr{p^2+p'^2-\qv\kv+k^2}{2m} - E_n \right) \times  \\
        && \left\{ k^2 + 2\pv\pp - 2\fr{(\kv\pp)(\kv\pv)}{k^2} -
            2i(\pv\times\kv,\Sv) - k^2\Sv^2 + (\kv\Sv)^2 \right\}. \non
\ea
In the second case all the elements of diagram 3 should be taken to
leading nonrelativistic approximation:
\ba\label{eq:MCC}
V_{MCC} &=& - \fr{\al^3}{2m^2} \int \fr{d^3 k}{(2\pi)^3}
            \int \fr{d^3 k'}{(2\pi)^3}\fr{(4\pi)^3}{k^4 (\qv-\kp)^2
            (\kp-\kv)^2 } \times  \\
        && \left\{ k^2 + 2\pv\pp - 2\fr{(\kv\pp)(\kv\pv)}{k^2} -
            2i(\pv\times\kv,\Sv) - k^2\Sv^2 + (\kv\Sv)^2 \right\}. \non
\ea
All the integrals in (\ref{eq:MC}) and (\ref{eq:MCC}) constitute
convolutions of the already known Fourier-transforms of powers of \r. In
this way we get for the first operator:
\ba
E^{(1)}_{MC} &=& \fr{\en}{2^5\cdot 3\cdot 5} \left\{ - 13 +
                 \fr{30}{n} - \fr{13}{n^2} \right. \non \\
              && \left. - \fr{3\Sv\lv}{5} \left( 29 + \fr{2}{3n^2} \right) +
                 2\fr{(\Sv\lv)^2 - 4\Sv^2}{5} \left( 13 - \fr{2}{n^2} \right)
                 \right\};
\ea
and for the second one:
\bq
E^{(1)}_{MCC} = \fr{\en}{2^5\cdot 3\cdot 5} \left\{ 15 \left( 1 -
                  \fr{1}{n} \right) + 9\Sv\lv - 2(\Sv\lv)^2 + 8\Sv^2 \right\}.
\eq

One more energy correction of the $\al^4 R_{\infty}$ order at the single
magnetic exchange is due to $Z$-type diagrams (see Fig.4). To leading
approximation one gets easily
\bq
V_{M-} = \fr{\al^2}{m^3} \int \fr{d^3 k}{(2\pi)^3} \fr{(4\pi)^2}{k^2 (\qv
           - \kv)^2} \left\{ \kv(\qv-\kv)( 1 - \Sv^2 ) +
         \kv\Sv\;(\qv-\kv)\Sv + \fr{1}{2} i(\kv\times\qv,\Sv) \right\}.
\eq
Standard calculations give now
\bq
E^{(1)}_{M-} = - \fr{\en}{2^2\cdot 5^2}
                 \left\{ \fr{5}{2} + \fr{3}{4}\Sv\lv - (\Sv\lv)^2 - \Sv^2
                 \right\}\left( 1 - \fr{2}{3n^2} \right).
\eq

\subsection{True radiative corrections}

Curiously enough, even the true radiative corrections of the $\al^4
R_{\infty}$ order to the $P$-levels energy can be presented in a
simple form practically without special calculation. They can be easily
demonstrated to be confined here to anomalous magnetic moment contribution
to the single magnetic exchange, i.e. to a trivial modification of the
usual Breit Hamiltonian. With this contribution the Pauli current becomes
\bq\label{eq:Pcm}
\vec{j}(\pp, \pv) \rightarrow \fr{1}{2m} ( \pp + \pv + ig\qv\times\sv ),
\eq
where
\bq
g = 1 + \fr{\al}{2\pi} - c \left( \fr{\al}{\pi} \right)^2 + \ldots, \;\;\;
c \approx 0.328
\eq
Let us note here that the anomalous magnetic moment contributions
to the first-order retardation effect and to diagram 2 cancel. It
corresponds to the absence of the order $v/c$ corrections to the Breit
magnetic exchange. In this way we get the following correction to the
$P$-level energy:
\bq
E^{(1)}_{rad} = - \fr{\en}{2^4\cdot 3\cdot 5} \fr{1}{\pi^2}
                \left\{ \left( \fr{3}{4} + 4c \right) \Sv\lv +
                ( 1 - 8c ) \left( \fr{3}{2}(\Sv\lv)^2 - \Sv^2 \right)\right\}.
\eq
Two its features are noteworthy. First, it vanishes in singlet states which
looks rather natural.  Second, it has an extra factor $1/\pi^2$ which
reflects its radiative origin as distinct from the relativistic origin of
other order $\al^4 R_{\infty}$ corrections to $P$-levels.

\subsection{Double magnetic exchange}

Let us consider now irreducible diagrams with two magnetic quanta. To our
approximation they are confined to the type presented in Fig.5. Their sum
reduces to
\ba\label{eq:MM}
V_{MM} &=& \fr{\al^2}{m^3} \int \fr{d^3 k}{(2\pi)^3} \fr{(4\pi)^2}{k^2 k'^2}
           \left\{ \pv\pp - 2\fr{(\kv\pv)(\kv\pp)}{k^2} +
           \fr{(\kv\pv)(\kv\kp)(\kv\pp)}{k^2 k'^2} - \fr{\kv\kp}{2}  \right.
           \non \\
       &&  + \left. i\Sv \left( \qv\times\pv - \kv\times\pv -
           \fr{\qv\times\kv\;(\kv\pp)}{k^2} \right) \right\},
\ea
where $\kp = \qv - \kv$. Now we substitute the products of operators in
the coordinate representation for the convolutions of their
Fourier-transforms and take the corresponding expectation values. In this
way we get
\bq
E^{(1)}_{MM} = \fr{\en}{2^4\cdot 3\cdot 5}
               \left\{ 10 \left( 1 - \fr{1}{n^2} \right) - 3\Sv\lv
               \left( 1 - \fr{2}{3n^2} \right) \right\}.
\eq

\section{Corrections of second order in the Breit \newline Hamiltonian}

Next class of the order $\alpha^4 R_{\infty}$ corrections originates from
the iteration of the usual Breit Hamiltonian $V$ of second order in
$v/c$.

Before proceeding to the calculation, let us note that the Breit potential
for the positronium (see, e.g., \cite{blp}, \S 84),
\ba \label{eq:VB}
V&=&- \fr{p^{4}}{4m^{3}}+\fr{\pi\al}{m^{2}}\delta(\r)
    - \fr{\al}{2m^{2}r}\left(p^{2}+(\n\pv)^2\right)
    + \fr{3\al}{2m^{2}r^{3}}\lv\Sv \nonumber \\
 && +\fr{3\al}{2m^{2}r^{3}}\left\{(\Sv \n )^2
    -\fr{1}{3}\vec{\cal S}^{2}\right\}
    + \fr{\pi\al}{m^{2}}\left(\fr{7}{3}\Sv^2-2\right)\delta(\r),
\ea
has nonvanishing matrix elements for $|\Delta L|=0,\,2$ only. A contribution of
virtual $F$-levels will be found later. And now we average the angular
dependence of (\ref{eq:VB}) over a $P$-state wave function:
\bq
3\left(\Sv\n\right)^2 - \Sv^2 \rightarrow -
\fr{3}{5} \left( 2\left(\Sv\lv\right)^2 + \Sv\lv - \fr{4}{3} \Sv^2 \right).
\eq
After this procedure the perturbation (\ref{eq:VB}) can be presented as
follows:
\ba
V &\rightarrow& \fr{m\al^4}{16} v, \\
v &=& \left( a\left\{h,\qu\right\} + b [ h, ip_r]
      + c \cb \right), \label{eq:v}
\ea
where
\bq\label{eq:koef}
a = -3, \;\;\;\; b = 1 + \kappa, \;\;\;\; c = -4 + \kappa;
\eq
\bq
\kappa = \fr{6\Sv\lv - 3(\Sv\lv)^2 + 2\Sv^2}{5};
\eq
$p_r = -i ( \de_r + 1/r )$ is the radial momentum, while
\[ h = \fr{p_r^2}{2} + \fr{1}{r^2} - \fr{1}{r} \]
is the unperturbed Hamiltonian for the radial motion with $L=1$ in the
Coulomb field.

According to the standard rules,
\ba
E_P^{(2)} &=& \langle VGV \rangle = \fr{m\al^6}{128} \langle vgv \rangle,
\label{eq:E2} \\
G &=& \sum_k{\displaystyle'}\fr{|kP\rangle\langle kP|}{E_n-E_k} =
      \fr{2}{m\al^2}g, \\
g &=& \sum_k{\displaystyle'}\fr{|kP\rangle\langle kP|}{\ep-{\cal E}_k}.
\ea
Since relativistic effects are included here into the perturbation,
the intermediate states $|kP\rangle$ are merely eigenfunctions of the
nonrelativistic Schr\"odinger equation in the Coulomb field.

The representation (\ref{eq:v}) enables us to find the energy correction
(\ref{eq:E2}) without recourse to the exact form of the Coulomb Green's
function $g$. Indeed,
\ba
\la vgv \ra &=& \fr{1}{2} \la \left( 2a\ep\qu - a\left[h,\qu\right]
      + b [ h, ip_r] + c \cb \right)gv \right. \nonumber \\
   && \left. + vg\left( 2a\ep\qu + a\left[h,\qu\right] + b [ h, ip_r]
      + c \cb \right) \ra \label{eq:1} \\
&=& -a\ep \left( 2\la v \ra + \de_\beta \la v \ra - \la \de_\beta v \ra
     \right) - \fr{a}{2} \la \{ v,\qu \} \ra \nonumber \\
   && + \fr{b}{2} \la [ip_r,v] \ra + c \left( \de_\gamma \la v \ra -
        \la \de_\gamma v \ra \right).\label{eq:2}
\ea
When passing from (\ref{eq:1}) to (\ref{eq:2}) we used the equation of
motion and presented the perturbations $1/r$ and $1/r^2$
as the result of $\beta$ and $\gamma$ variations in the modified
Hamiltonian,
\bq
h(\beta,\gamma) = \fr{p_r^2}{2} + \fr{\gamma}{r^2} - \fr{\beta}{r}.
\eq
Derivatives in (\ref{eq:2}) are taken at $\beta=\gamma=1$.

Substituting (\ref{eq:v}) into (\ref{eq:2}) provides us with the expression
containing just the mean values of $1/r^k$ (see (\ref{eq:means})).

We collect now coefficients at the powers of $1/n$ to obtain:
\ba
E_P^{(2)} &=& \fr{m\al^6}{128n^3} \left\{ -\fr{3a^2+14ab+13b^2}{15} -
        \fr{2c(2c+9a+9b)}{27} \right. \nonumber \\
   && \left. - \fr{2c^2}{3n} + \fr{2}{3n^2}\left( \fr{11a^2 + 13ab +
      6b^2}{5} + 4ac \right) - \fr{5a^2}{2n^3} \right\}.
\ea
Finally, substituting the expressions for $a,b,c$ (\ref{eq:koef}), we get
the contribution of intermediate $P$ states to the iteration of the Breit
Hamiltonian:
\bq
E_P^{(2)}=\en
          \left\{ - \fr{1022-844\kappa+227\kappa^2}{2^7\cdot 3^3\cdot 5}
          -\fr{(4-\kappa)^2}{2^6\cdot 3 \cdot n}
          + \fr{102-29\kappa+2\kappa^2}{2^6\cdot 5 \cdot n^2} -
          \fr{45}{2^8 \cdot n^3} \right\}.
\eq

Transitions to intermediate $F$ states are driven by the operator
$(\Sv\n)^2$ contained in (\ref{eq:VB}). Due to the conservation of the
total angular momentum such transitions are possible in the states
with $J=2$ only. The matrix element squared of the operator $(\Sv\n)^2$ is
easily calculated by means of closure:
\ba
|\langle P_2 |(\Sv\n)^2 | F_2 \rangle |^2 &=&
\langle P_2 |(\Sv\n)^4 | P_2 \rangle -
\langle P_2 |(\Sv\n)^2 | P_2 \rangle^2 \nonumber \\
&=& \langle P_2 |(\Sv\n)^2 | P_2 \rangle -
\langle P_2 |(\Sv\n)^2 | P_2 \rangle^2 \;=\fr{6}{25}.
\ea
We have used here the identity $(\Sv\n)^3 | P \rangle = \Sv\n | P \rangle$.

To calculate the radial part of the correction, proportional to
\bq
\la \fr{1}{r^3}\sum_k\fr{|kF\rangle\langle kF|}{\ep-{\cal E}_k}\fr{1}{r^3}
\ra,
\eq
we present the operator $1/r^3$ in the form
\ba
\fr{1}{r^3} &=& \fr{1}{18} \left[h\left(-ip_r-\fr{5}{r}+\fr{1}{5}\right) -
\left(-ip_r-\fr{5}{r}+\fr{1}{5}\right)h_F \right] \\
&=& \fr{1}{18} \left[h_F\left(-ip_r+\fr{5}{r}-\fr{1}{5}\right) -
\left(-ip_r+\fr{5}{r}-\fr{1}{5}\right)h \right],
\ea
where
\bq
 h_F = \fr{p_r^2}{2} + \fr{6}{r^2} - \fr{1}{r}
\eq
is the unperturbed Hamiltonian for the radial motion with $L=3$ in the
Coulomb field. The energy correction is again expressed in terms of
$\langle r^{-k} \rangle$:
\ba
E_F^{(2)} &=& m\al^6 \fr{3}{2^8\cdot 5^2} \la \fr{2}{5r^3} - \fr{7}{r^4} \ra
\nonumber \\
&=& - \fr{\en}{2^6\cdot 5^3} \left( 10 - \fr{7}{n^2} \right).
\ea

\section{Numerical results}

Let us summarize the numerical values of the ${\alpha}^4 R_{\infty}$
corrections to the energies of positronium $2 P$ levels. They are

\ \ \ 0.06 MHz for $2^1P_1$,

\ \ \ 0.08 MHz for $2^3P_2$,

\ \ \ 0.025 MHz for $2^3P_1$,

\ -- 0.58 MHz for $2^3P_0$.

We wish to emphasize here that the last correction is quite comparable
in magnitude to the corresponding logarithmic correction (of order
$\al^4 \log \al \; R_{\infty}$) to the positronium $2S$-levels. The latter
constitutes for instance 0.96 MHz for the $2 ^3 S_1$ state \cite{fel,kmy}.
Therefore, there is no special reasons to expect that the nonlogarithmic
corrections (of order $\al^4 R_{\infty}$) to the positronium $S$ levels are
small. It makes their calculation quite an actual problem.

Including ${\alpha}^2 R_{\infty}$ and ${\alpha}^3 R_{\infty}$ terms we
obtain the fine-structure intervals between $2^{2S+1}P_J$ levels. In the
Table 1 these theoretical values are compared with the transition
frequencies extracted from the results of two recent experiments
\cite{hcr,chl,hage,hag2} (all systematic and statistical errors are added
quadratically when extracting the experimental numbers).
\bigskip

{\bf Acknowledgements}

We are grateful to R. Conti, D. Hagena and G. Werth for the communication of
their results prior to publication and for the explanations how to treat the
errors when extracting the frequency differences.  We are grateful also to
J. Sapirstein who attracted our attention to Ref \cite{dk}. The work was
supported by the Russian Fund for Fundamental Research. One of us (I.B.
Kh.) wishes to thank Institute for Nuclear Theory, University of Washington,
Seattle, for the kind hospitality during the stay when major part of his
work has been done.

\newpage

Table 1:   Fine-structure intervals between $2^{2S+1}P_J$ levels (in MHz).
\begin{center}
\begin{tabular}{|c|c|c|c|c|} \hline
   & \multicolumn{2}{c|}{Experiments} & \multicolumn{2}{c|}{Theory}\\ \hline
   & Michigan \cite{hcr,chl} & Mainz \cite{hage,hag2} & Total &
$\alpha^4 R_{\infty}$ contribution\\ \hline
$E\left(^3P_2\right)-E\left(^3P_0\right)$ & 9884.5$\pm$10.5 & 9875.27
$\pm$4.44  & 9871.54 & 0.66 \\
$E\left(^3P_1\right)-E\left(^3P_0\right)$ & 5502.8$\pm$10.9 & 5487.23
$\pm$4.50  & 5485.84 & 0.60 \\
$E\left(^1P_1\right)-E\left(^3P_0\right)$ & 7323.1$\pm$16.5 & 7319.65
$\pm$7.64  & 7312.88 & 0.64 \\ \hline
\end{tabular}
\end{center}

\newpage

{\bf Figure captions}

\noindent {\it Fig.1.} Z-type double-Coulomb exchange

\noindent {\it Fig.2.} Single-magnetic-single-Coulomb exchange

\noindent {\it Fig.3.} Single-magnetic-double-Coulomb exchange

\noindent {\it Fig.4.} Z-type single-magnetic-single-Coulomb exchange

\noindent {\it Fig.5.} Double-magnetic exchange

\newpage

\begin{figure}
 \begin{picture}(460,180)
  \put(115,40){\begin{picture}(230,140)
              \thicklines
              \put(10,110){\line(1,0){140}}
              \put(210,70){\vector(-1,0){120}}
              \put(90,70){\vector(3,2){60}}
              \put(210,20){\vector(-1,0){60}}
              \put(150,20){\vector(-1,0){60}}
              \put(90,20){\line(-1,0){80}}
              \multiput(90,20)(0,20){3}{\line(0,1){10}}
              \multiput(150,20)(0,20){5}{\line(0,1){10}}
              \put(90,0){Fig.1}
            \end{picture}}
 \end{picture}
\end{figure}

\begin{figure}
 \begin{picture}(460,180)
  \put(0,40){\begin{picture}(230,140)
             \multiput(200,30)(-40,20){5}{\oval(40,20)[bl]}
             \multiput(160,30)(-40,20){4}{\oval(40,20)[tr]}
             \thicklines
             \multiput(210,20)(-90,90){2}{\vector(-1,0){100}}
             \multiput(210,20)(0,90){2}{\line(-1,0){200}}
             \multiput(110,20)(0,20){5}{\line(0,1){10}}
             \put(90,0){Fig.2}
             \end{picture}}
  \put(230,40){\begin{picture}(260,140)
              \multiput(200,30)(-40,20){5}{\oval(40,20)[bl]}
              \multiput(160,30)(-40,20){4}{\oval(40,20)[tr]}
              \thicklines
              \multiput(200,20)(-60,90){2}{\vector(-1,0){60}}
              \multiput(140,20)(-60,90){2}{\vector(-1,0){60}}
              \multiput(10,20)(0,90){2}{\line(1,0){200}}
              \multiput(80,20)(0,20){5}{\line(0,1){10}}
              \multiput(140,20)(0,20){5}{\line(0,1){10}}
              \put(90,0){Fig.3}
              \end{picture}}
 \end{picture}
\end{figure}

\begin{figure}
 \begin{picture}(460,180)
  \put(0,40){\begin{picture}(230,140)
            \multiput(160,30)(-20,20){4}{\oval(20,20)[bl]}
            \multiput(140,30)(-20,20){3}{\oval(20,20)[tr]}
            \thicklines
            \multiput(190,20)(0,90){2}{\vector(-1,0){160}}
            \put(190,20){\vector(-1,0){30}}
            \put(30,110){\vector(3,-1){60}}
            \multiput(90,20)(0,70){2}{\line(-1,0){80}}
            \multiput(30,20)(0,20){5}{\line(0,1){10}}
            \put(90,0){Fig.4}
            \end{picture}}
  \put(230,40){\begin{picture}(260,140)
               \multiput(190,30)(-20,20){4}{\oval(20,20)[bl]}
               \multiput(170,30)(-20,20){3}{\oval(20,20)[tr]}
               \multiput(60,60)(20,40){2}{\oval(20,60)[br]}
               \multiput(60,20)(20,40){3}{\oval(20,20)[tl]}
               \thicklines
               \multiput(160,20)(50,90){2}{\vector(-1,0){110}}
               \put(210,20){\vector(-1,0){20}}
               \put(100,110){\vector(1,-1){20}}
               \multiput(70,20)(120,0){2}{\line(-1,0){40}}
               \put(30,90){\line(1,0){90}}
               \put(90,0){Fig.5}
               \end{picture}}
 \end{picture}
\end{figure}

\end{document}